# Testing the Large-scale Environments of Cool-core and Non-cool-core Clusters with Clustering Bias

Elinor Medezinski[1], Nicholas Battaglia[1], Jean Coupon[2], Renyue Cen[1], Massimo Gaspari[1,3],
Michael A. Strauss[1], and David N. Spergel[1]
[1] Department of Astrophysical Sciences, 4 Ivy Lane, Princeton, NJ 08544, USA; elinorm@astro.princeton.edu
[2] Department of Astronomy, University of Geneva, ch. dEcogia 16, CH-1290 Versoix, Switzerland

## Abstract

There are well-observed differences between cool-core (CC) and non-cool-core (NCC) clusters, but the origin of this distinction is still largely unknown. Competing theories can be divided into *internal* (inside-out), in which internal physical processes transform or maintain the NCC phase, and *external* (outside-in), in which the cluster type is determined by its initial conditions, which in turn leads to different formation histories (i.e., assembly bias). We propose a new method that uses the relative assembly bias of CC to NCC clusters, as determined via the two-point cluster-galaxy cross-correlation function (CCF), to test whether formation history plays a role in determining their nature. We apply our method to 48 ACCEPT clusters, which have well resolved central entropies, and cross-correlate with the SDSS-III/BOSS LOWZ galaxy catalog. We find that the relative bias of NCC over CC clusters is $b = 1.42 \pm 0.35$ ($1.6\sigma$ different from unity). Our measurement is limited by the small number of clusters with core entropy information within the BOSS footprint, 14 CC and 34 NCC clusters. Future compilations of X-ray cluster samples, combined with deep all-sky redshift surveys, will be able to better constrain the relative assembly bias of CC and NCC clusters and determine the origin of the bimodality.

## 1. Introduction

Clusters of galaxies grow hierarchically through mergers and accretion of galaxies and groups of galaxies. The gas that falls onto a cluster gravitationally shocks and heats to the observed virial temperature, $T \sim 10^8$ K. In a simplified gravitationally governed smooth accretion model, clusters should have self-similar entropy profiles (Voit 2005). A decade of high-resolution *Chandra* X-ray observations of cluster gas has unveiled relatively self-similar scaled entropy profiles at virial radius scales ($\sim$1–2 Mpc/$h$), but the central ($\lesssim$0.1–0.2 Mpc/$h$) entropies differ considerably. Some clusters show relaxed, cuspy cores with high metallicity and low central temperature and entropy, and are thus named cool core (CC; Molendi & Pizzolato 2001), whereas others show a more disturbed core with flatter central density and high core entropy, dubbed non-cool core (NCC). Classification schemes vary because CC are hard to define and quantify (Hudson et al. 2010), but the central entropy has been shown to be the best classifier (Cavagnolo et al. 2009).

In a purely radiative cooling scenario, the plasma in the cores of clusters is expected to condense in less than a Gyr, with cooling rates up to $10^3\,M_\odot\,\mathrm{yr}^{-1}$. This would lead to dramatic star formation rates and very peaked X-ray surface brightness profiles. The observed CC show much more gentle cooling rates, 1%–10% of the pure cooling flow value (for a review, see Gaspari 2015 and references therein). In contrast to NCC clusters, observed CCs typically have central temperatures that are two to three times smaller than the virial value, where entropy starts to flatten (McNamara & Nulsen 2007, 2012). Mechanical active galactic nucleus (AGN) feedback is the current best model to explain the quenching of pure cooling flows, although other forms of heating may contribute (e.g., thermal conduction and cosmic rays; McNamara & Nulsen 2007, 2012 for reviews). Cool cores are indeed correlated with the presence of X-ray cavities inflated by AGN outflows (e.g., Hlavacek-Larrondo et al. 2015), low central entropy/cooling times (Cavagnolo et al. 2008), large H$\alpha$ luminosity (e.g., Voit & Donahue 2015), and multiphase gas down to the molecular regime (e.g., Tremblay et al. 2016). Such residual cooling gas is the main triggering mechanism of the AGN, leading to a tight self-regulated loop between CC condensation and AGN-feedback energy (e.g., Gaspari et al. 2017 and references therein).

Hydrodynamic simulations on scales of both individual halos and cosmological volumes have improved through the introduction of sub-grid physics models for various forms of feedback. However, simulating the relevant scales for these feedback processes is still a prodigious numerical challenge (Borgani & Kravtsov 2011, and references therein). A considerable amount of numerical effort has been spent on understanding the suppression of cooling flows via feedback mechanisms (e.g., Heinz et al. 2006; Sijacki & Springel 2006; Battaglia et al. 2010; Dubois et al. 2010; Gaspari et al. 2011; McCarthy et al. 2011; Martizzi et al. 2012; Li & Bryan 2014; Li et al. 2015; Steinborn et al. 2015). Recent simulations have been able to create a diverse sample of CC and NCC clusters (Rasia et al. 2015; Hahn et al. 2015). However, the cosmological hydrodynamic simulations are still significantly limited by the poor resolution within the CC region, which is crucial to properly track the AGN heating distribution and turbulent mixing properties, as shown in previous high-resolution simulations of isolated clusters (e.g., Gaspari et al. 2011, 2014a). The question remains, what is the physical origin of the differences between CC and NCC clusters? Were they created under different physical conditions to begin with,

---

[3] Einstein and Spitzer Fellow.

or are NCCs simply disturbed CCs that have not yet had the time to cool back down?

There are two scenarios for how CCs and NCCs form and evolve—*external* or *internal*. In the external scenario, outside factors such as the large-scale structure would play a central role in determining the fate of clusters ab-initio. In this model, NCC would typically be found in denser environments, whereby merger activity can pre-heat the clusters to higher levels (∼300 Kev cm$^2$; McCarthy et al. 2011). CC, on the other hand, would tend to form in isolation. In the internal scenario, only the immediate (inside ∼1 Mpc/$h$) environment acts to transform CC to NCC. Here, breaking the tight self-regulated AGN-feedback loop results in overheating the core and raising the central cooling time well above the Hubble time (Gaspari et al. 2014b). Since in this case AGN heating should be unrealistically strong, it is more plausible that infalling substructures within the cluster (Sanderson et al. 2006, 2009; Leccardi et al. 2010; Rossetti & Molendi 2010; Rossetti et al. 2011; Eckert et al. 2014) are responsible for breaking the loop, thus inducing the NCC state.

To test how much large-scale environment plays a role in shaping these cluster cores, one can exploit galaxy clustering. The clustering of collapsed halos is enhanced relative to the dark matter distribution, an effect known as bias (Kaiser 1984; Efstathiou et al. 1988; Cole & Kaiser 1989; Bond et al. 1991; Mo & White 1996; Sheth & Tormen 1999). This bias depends mostly on halo mass, making galaxy clusters highly biased (Bahcall & Soneira 1983; Kaiser 1984). However, numerical simulations show there is an additional but weaker dependence on the formation histories of the halos, an effect that is referred to as *assembly bias* (Gao et al. 2005; Wechsler et al. 2006; Gao & White 2007; Jing et al. 2007; Wetzel et al. 2007; Angulo et al. 2008). On cluster-mass scales, one predicts that late-forming (low-concentration) objects of a given mass are more clustered (Wechsler et al. 2006; Jing et al. 2007; Wang et al. 2007; Zentner 2007; Dalal et al. 2008), but the effect is expected to be even weaker than on galaxy scales (Gao et al. 2005). Assembly bias has been difficult to demonstrate conclusively in observations, because it requires identifying samples that have similar halo mass, but differ in assembly histories. A handful of observational studies have tried to measure assembly bias in the regime of groups (Yang et al. 2006; Wang et al. 2013; Lacerna et al. 2014) and clusters (Miyatake et al. 2016; More et al. 2016). However, Lin et al. (2016) argue that the claimed detections of assembly bias on group scales could be attributed to samples of different halo mass or contamination by satellite galaxies rather than assembly bias. So far, attempts to detect assembly bias have divided cluster samples according to halo concentration as a formation epoch proxy (Miyatake et al. 2016; More et al. 2016), as higher concentration is linked with earlier formation in numerical simulations (Duffy et al. 2008; Bhattacharya et al. 2011). No study has yet attempted to detect it for halos of different X-ray properties, such as entropy.

In this paper, we use spatial cross-correlations between galaxies and galaxy clusters to explore the large-scale environments, and hence, the assembly bias of CC versus NCC clusters. We use a statistical sample of clusters with information on their entropic core state from the ACCEPT compilation (Cavagnolo et al. 2009). This paper is organized as follows. In Section 2 we present the observational data set used. In Section 3 we lay out our CCF methodology and show how to derive a relative bias. In Section 4 we present our results, and in Section 5 we forecast the improvement to our results with larger cluster samples. We summarize and conclude in Section 6. Throughout the paper, we adopt a ΛCDM cosmological model, where $\Omega_m = 0.27$, $\Omega_\Lambda = 0.73$, and $h = H_0/100$ km s$^{-1}$ Mpc$^{-1} = 1$.

## 2. Data

We compare the clustering of two subsets of galaxy clusters: CC and NCC. The auto-correlation will be very noisy because our cluster samples are small (≲30 clusters in each, see below), so instead we perform a cross-correlation of each cluster sample with a parent galaxy sample. In this section we present the cluster and galaxy samples.

### 2.1. ACCEPT Cluster Samples

The largest publicly available homogeneous compilation of X-ray clusters with high-resolution radial entropy profiles[4] was presented in Cavagnolo et al. (2009),[5] known as the Archive of *Chandra* Cluster Entropy Profile Tables (ACCEPT). This sample consists of the 241 clusters with enough counts for a reliable entropy determination that have been observed with the *Chandra* X-ray telescope (each cluster temperature profile listed has at least three concentric radial annular bins containing a minimum of 2500 source counts each), and were in the archive as of 2008.[6]

Cavagnolo et al. (2009) found that most ICM entropy profiles are well fitted by a power-law model at large cluster radii and approach a constant value at small radii (≲100 kpc), $K_0$. This entropy floor quantifies the typical excess of the core entropy relative to a strict power law, and was shown to provide acceptable fits for >90% of the ACCEPT clusters. However, for clusters with low surface brightness X-ray emission at the core, the central entropy is not as well constrained. To avoid this extrapolation, we opt to directly measure the entropy at 20 kpc, where all the clusters have resolved entropy information. We use the full entropy profiles from the ACCEPT database and interpolate the value at 20 kpc. In Figure 1 (gray points) we compare Cavagnolo's $K_0$ with $K_{20}$, and find good correlation, especially at large central entropies. We note that the crude threshold of $K_0 \approx 40$ keV cm$^2$ used in Cavagnolo et al. corresponds roughly to $K_{20} = 60$ keV cm$^2$ (magenta dotted line). This entropy translates to a typical cooling time of 1 Gyr (see Equation (9) in Cavagnolo et al. 2009), below which clusters are expected to host a strong CC (Hudson et al. 2010). We chose this value as a boundary to divide the ACCEPT sample into two cluster subsamples—those with lower $K_{20}$ (shorter cooling times) are considered CC (blue points; after applying spatial and redshift cuts described below) and those with larger $K_{20}$ (longer cooling times) are considered NCC (red points). ACCEPT clusters appear evenly distributed over the sky (Figure 2; CC in blue and NCC in red), and span redshifts in the range of $0.05 < z < 1.1$ (see the redshift distribution in Figure 3).

---

[4] There is a larger compilation of X-ray clusters drawn from *Chandra*—the *ROSAT* All-Sky Survey (RASS) and XMM-Newton (Piffaretti et al. 2011)—but they lack the high spatial resolution that *Chandra* provides to allow for homogeneous CC/NCC classification.
[5] http://www.pa.msu.edu/astro/MC2/accept/
[6] Many clusters have since been observed with *Chandra*, and an ACCEPT-2 compilation (M. Donahue 2017, private communication) of entropy profiles is being prepared and will be utilized in a future paper.

### 2.2. LOWZ Galaxy Sample

For the large parent sample we use the LOWZ spectroscopic-redshift galaxy catalog[7] (Reid et al. 2016), which is drawn from the Baryon Oscillation Spectroscopic Survey (BOSS; Dawson et al. 2013). BOSS is part of Sloan Digital Sky Survey (SDSS; Eisenstein et al. 2011) III project Data Release 12 (DR12; Alam et al. 2015). The LOWZ sample is designed to extend the SDSS-I/II (York et al. 2000) Cut I luminous red galaxy (LRG) sample (Eisenstein et al. 2001) to $z \approx 0.4$ and fainter luminosities to increase the number density of LRGs by roughly a factor of 3. In DR12 the survey is complete—covering both the north Galactic pole (NGC) and south Galactic pole (SGC)—with a total effective area of 8337 deg$^2$, and containing 463,044 galaxies with spectroscopic redshifts (Reid et al. 2016). The sample distribution over the sky is presented in Figure 2 (gray) and its redshift distribution is presented in Figure 3 (gray). In recent analyses of LOWZ (Cuesta et al. 2016; Reid et al. 2016), the redshift range used was limited to $0.15 < z < 0.43$, resulting in a total of 361,762 galaxies. The sample is close to volume-limited (constant space density at $\sim 3 \times 10^4\, h^3$ Mpc$^3$) over the redshift range $0.2 < z < 0.4$ (Reid et al. 2016). We describe below the redshift limits we choose to maximize the use of the cluster sample.

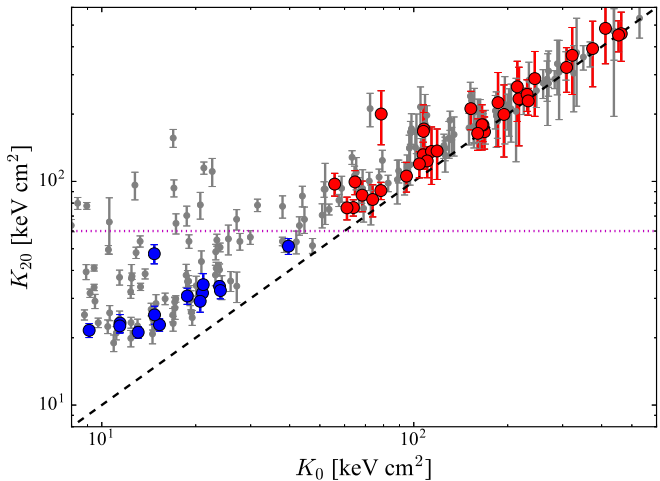

**Figure 1.** Different central entropy definitions compared—$K_0$ from the fitted profile of Cavagnolo et al. (2009) vs. $K_{20}$—the entropy measured at $r = 20$ kpc for all ACCEPT clusters (gray). The magenta dotted line marks the chosen boundary, $K_{20} = 60$ keV cm$^2$, that separates CC (blue) from NCC (red) clusters.

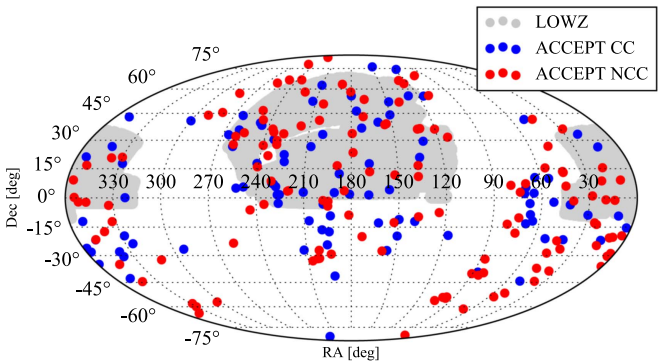

**Figure 2.** Sky distribution of the LOWZ galaxy sample (gray) and ACCEPT CC (blue) and NCC (red) clusters.

### 2.3. Catalog Spatial and Redshift Limits

We limit our cluster and galaxy catalogs such that they span the same sky area and redshift range for our cross-correlation measurement. In Figure 2 we present the spatial distribution of all LOWZ galaxies (gray) and ACCEPT clusters (red+blue points); 102 of the 241 ACCEPT clusters are found within the BOSS NGC and SGC footprints. We present the redshift distribution of the LOWZ galaxies (gray) and ACCEPT clusters found in BOSS footprint (CC in blue and NCC in red) in Figure 3. The LOWZ sample is typically analyzed within the $0.15 < z < 0.43$ range, where its selection function is reasonably uniform and well understood. The completeness of the LOWZ sample appears robust to $z = 0.1$, so we choose here to expand the low-redshift limit to $z > 0.1$ to overlap with low-redshift ACCEPT clusters. We note that a similar lower limit has been used for other cluster sample analyses using SDSS, namely the redMaPPer cluster catalog (Rykoff et al. 2014, 2016; Miyatake et al. 2016). Within the BOSS footprint and redshift range, $0.1 < z < 0.43$, there are 400,176 LOWZ galaxies and 57 ACCEPT clusters, out of which 23 are CC and 34 are NCC.

### 2.4. Mass Difference

To leading order, halo bias depends on mass, thus, it is important to ensure that the two cluster samples have the same mean mass before any statement on second order effects such as assembly bias can be made. We match the ACCEPT clusters within the BOSS footprint with the Planck SZ cluster sample (Planck Collaboration et al. 2016) and use the Planck SZ masses. We match 14 of the 23 CC clusters, and all 34 NCC clusters. The remaining clusters presumably fall below the Planck mass detection limit. We find that the ratio of mean masses as determined from the Planck SZ mass is $\langle M_{\rm SZ,NCC} \rangle / \langle M_{\rm SZ,CC} \rangle = 1.035 \pm 0.032$ for this subsample.

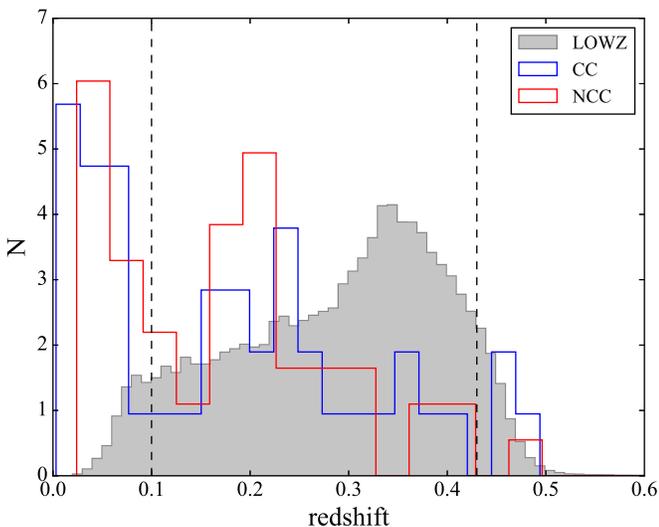

**Figure 3.** Redshift distribution of different samples: LOWZ galaxies (gray) and ACCEPT CC (blue) and NCC (red) clusters within the BOSS FOV. The dashed vertical lines show our chosen redshift limits. Only 57 clusters are within the redshift range and the BOSS footprint. The histograms are normalized for easier comparison. Within the redshift range chosen, the CC and NCC redshift distributions are statistically indistinguishable.

---
[7] https://data.sdss.org/sas/dr12/boss/lss/

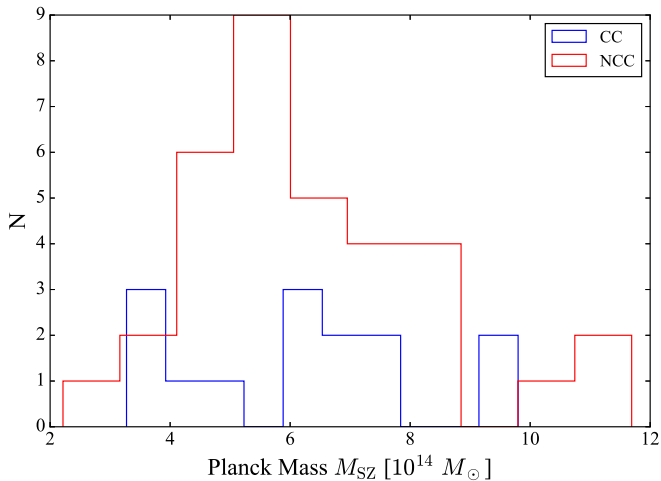

**Figure 4.** Mass distribution of 14 CC (blue) and 34 NCC (red) ACCEPT clusters within the BOSS FOV that are matched with Planck clusters.

The mass distributions of the two cluster samples are presented in Figure 4. A Kolmogorov–Smirnov (KS) test confirms that the two subsample masses are likely drawn from the same mass distribution, with $D_{\rm KS} = 0.2$ and a $p$-value of $p = 0.9$.

### 3. Cross-correlation Functions and Relative Bias

The two-point correlation function is a measure of how spatially clustered two populations are. One can formulate the probability above random of finding a galaxy in a volume element $dV$ at a distance separation $r$ from another galaxy as

$$dP(r) = n[1 + \xi(r)]dV \quad (1)$$

where $n$ is the mean number density and $\xi(r)$ is the two-point correlation function (Peebles 1980). If objects are distributed uniformly in space $\xi(r) = 0$, whereas $\xi(r) > 0$ indicates clustering. The two-point CCF between galaxies and clusters in our case relates to the probability as

$$dP(r) = n_g n_c [1 + \xi_{gc}(r)]dV_g dV_c, \quad (2)$$

where $n_g, n_c$ are the galaxy and cluster number densities, respectively. In the linear bias approximation, the galaxy auto-correlation function is related to the underlying dark matter by $\xi_{gg}(r) = b_g^2 \xi_{\rm DM}(r)$. For the galaxy-cluster CCF, this relation is

$$\xi_{gc}(r) = b_g b_c \xi_{\rm DM}(r). \quad (3)$$

Since we correlate each of the two cluster samples (CC, NCC) with the same galaxy sample, the ratio of these cross-correlations simply traces the *relative* bias of NCC with respect to CC clusters,

$$b \equiv b_{\rm NCC}/b_{\rm CC} = \frac{\xi_{\rm NCC}}{\xi_{\rm CC}}. \quad (4)$$

To estimate the 3D CCF, we count the number of galaxy-cluster pairs, $D_g D_c(r)$, in bins of comoving separation, $r$, and compare with corresponding pair counts drawn from equivalent random galaxy and cluster catalogs, $R_g$, $R_c$, respectively, in each bin. We make use of the modified Landy & Szalay (1993; hereafter LS) estimator to calculate the CCF of these pairs,

$$\xi_{\rm gc}(r) = \frac{D_g D_c(r) - D_g R_c(r) - D_c R_g(r) + R_c R_g(r)}{R_c R_g(r)}, \quad (5)$$

where each data and random catalog are normalized by their number density.

The selection function of clusters having *Chandra* observations is not defined, thus a cluster random catalog, $R_c$, is also impossible to construct. We instead simply match the sky and redshift distribution of the galaxy sample, as described above, and use the galaxy random catalog provided by Reid et al. (2016) for $R_c$. It is only important that the two cluster samples be drawn from the same distribution, because we are interested in the *ratio* between CC and NCC clustering. Figure 3 shows that the redshift distributions of the two cluster subsamples are similar in our chosen redshift range. A KS test supports that the two samples are drawn from the same redshift distributions, with $D_{\rm KS} = 0.2$ and a high $p$-value, $p = 0.6$.

### 4. Results

We make use of the public code SWOT[8] (Coupon et al. 2012) to calculate the CCF of ACCEPT clusters with LOWZ galaxies in six logarithmic bins spanning 3–80 Mpc/$h$. We note that above 3 Mpc/$h$ separation we are safely at the two-halo regime, and thus avoid cross-correlating clusters with their own satellite galaxies. Finger-of-god effects are also not expected to be significant in this regime, as will be evident by the large errors on the resulting CCFs presented below. SWOT uses a descending k-d tree approach to cross-correlate catalogs, and has a lower opening angle threshold (OA) below which k-d trees are not further descended and large-scale distances are approximated to speed up processing time. We set OA = 0.03 radians, but find this has no effect on our results in the examined range, $r < 80$ Mpc/$h$. The resulting CCFs are presented in Figure 5 (left) for the CC subsample (blue circles) and for the NCC subsample (red triangles). The bias, given as the ratio of the two, is presented in the bottom panel. We compare LS with the Davis & Peebles (1983) estimator, and find that the results are identical.

By construction, adjacent bins of the correlation function are highly correlated, especially at large separations. Therefore, Poisson errors underestimate the true variance on the large scales examined here. There are many different approaches in the literature to account for this covariance. One popular way is to divide the sky into equal subregions and derive the covariance using the jackknife method (Mountrichas et al. 2009, 2016; Miyatake et al. 2016). Another is to perform the calculation over many mock simulations that mimic the samples at hand (see, e.g., Blake et al. 2006; Knobel et al. 2012). Mock simulations are preferred, as the jackknife value distribution is not Gaussian for a small sample like the one presented here. Furthermore, the largest separation that can be probed is limited by the jackknife region size, because large-scale modes are not probed by the smaller jackknife box (Norberg et al. 2009). Thus the errors on all quantities are derived using simulations, where we cross-correlate samples of mock galaxies and clusters of sizes comparable to the data at hand (for full details of the simulations and error analysis, see Appendix).

---

[8] http://jeancoupon.com/swot

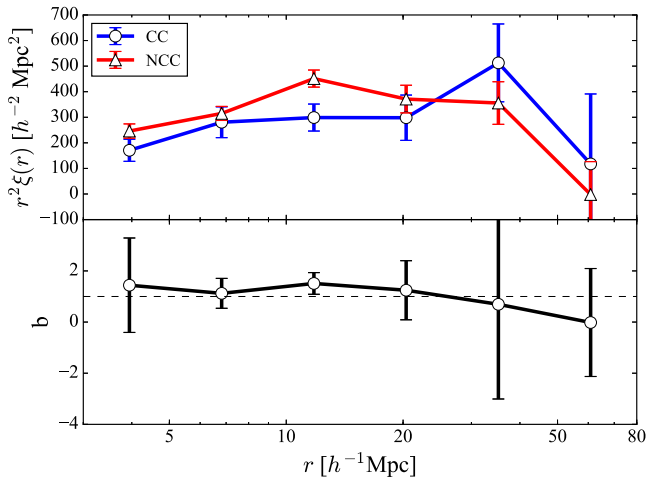
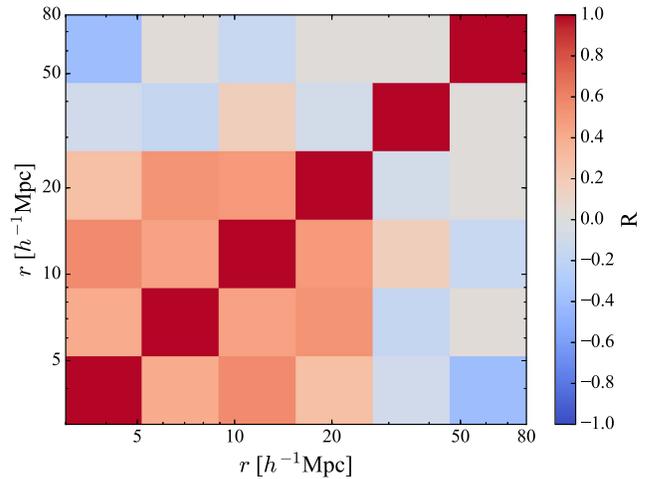

**Figure 5.** Left: CCFs between LOWZ galaxies and two cluster subsamples: CC clusters (blue curve+circles) and NCC clusters (red curve+triangles). Bottom panel shows the ratio of NCC to CC correlations, which gives the relative bias, $b$. Errors are drawn from the scatter of 50 mock MICE CCF and bias measurements (same errors as in Figure 7, left; see Appendix for details). Right: Correlation coefficients of the bias determined from the simulation's covariance matrix.

As is evident from the size of the errors in the left panel of Figure 5, the ratio of CCFs at large scales ($\gtrsim 25\,\mathrm{Mpc}/h$) is noisy because it is dividing two small quantities—the CCF at these scales approaches zero. When considering the full covariance derived from the simulations (presented in the right panel of Figure 5), the mean relative bias is $\langle b \rangle = 1.42 \pm 0.35$, with a significance of $1.6\sigma$ relative to $b = 1$. In short, within the large uncertainty, we currently do not find a significant difference in the clustering around NCC and CC clusters.

## 5. Forecast

We find that the average bias is determined at the $\Delta b/b \approx 0.35/1.42 = 25\%$ level. Although cross-correlating with a large galaxy sample allowed us to analyze a small cluster sample ($\langle N_{\mathrm{cl}} \rangle \sim 24$), our analysis is still mostly limited by the small number of clusters with central entropy measurements. Here we forecast how a larger cluster sample will help improve this statistic. We perform the same mock analysis using simulations as described in the Appendix for increasing numbers of mock clusters, $N_{\mathrm{cl}} = 100, 250, 500$, in each subsample. In Figure 6 we plot the size of bias errors, derived from the covariance of 50 mock simulations, as a function of cluster sample size. We plot this as a function of scale, $r$. The expected slope of this relation according to Poisson statistics, $-1/2$, is overlaid to guide the eye (black dashed–dotted line). For $r \lesssim 35\,\mathrm{Mpc}/h$, the errors roughly follow the Poisson expectation, but they do not at $70\,\mathrm{Mpc}/h$. Following this scaling, a future sample of 500 clusters in each CC/NCC subsample will lead to a constraint on the bias that is $\lesssim 5\%$, and best probed over scales $r \lesssim 20\,\mathrm{Mpc}/h$.

The level of assembly bias we find is in statistical agreement with the level found by Miyatake et al. (2016), $b = 1.41 \pm 0.09$. The main differences between these two measurements are the cluster samples and how those samples are subdivided. Miyatake et al. (2016) use cluster-galaxy member concentration as a proxy for their formation histories. Simulations predict the level of bias to be $\sim 1.2$ (Wechsler et al. 2006; More et al. 2016). Assuming this level of bias, with $\sim 500$ clusters in each subsample we can make a significant detection of assembly bias using our method at the $3\sigma$ level.

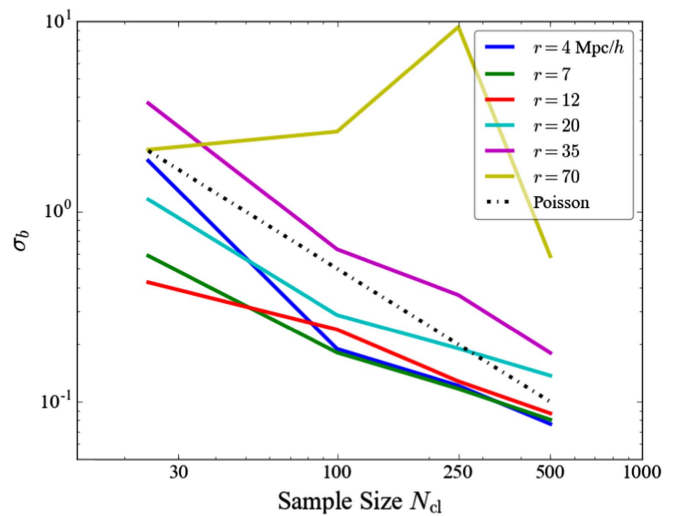

**Figure 6.** Error in the bias derived from simulations (solid lines) as a function of mock cluster sample size, plotted for each separation scale separately (different colors), as noted in the legend. The expected Poisson error scaling, $\sqrt{N}$, is shown as the dashed–dotted line (with arbitrary amplitude).

## 6. Discussion and Conclusions

In this paper, we presented a methodology to explore the origin of the CC-NCC dichotomy. We compared the clustering of BOSS/LOWZ galaxies around CC and NCC cluster samples. By comparing these CCFs, we constrain the relative assembly bias of NCC with respect to CC clusters to be $\langle b \rangle = 1.42 \pm 0.35$, which is only $1.6\sigma$ above unity (a null detection). Limited by the small number of clusters in our subsamples (14 CC and 34 NCC), we do not detect a significant difference between the large-scale environments of CC and NCC clusters.

The main limitation of the current proposed method is the number of clusters with resolved X-ray cores available. We note that our study was done with a sample of *Chandra* X-ray clusters compiled nearly a decade ago. Since then, many follow-up *Chandra* cluster observations have been carried out, in particular targeting the Planck $z < 0.35$ clusters (Jones 2012). An updated compilation of cluster entropy profiles using deprojected temperatures is currently being prepared, and will increase the number of clusters from ∼200 (ACCEPT) to at least ∼500 (ACCEPT-2; Baldi et al. 2014). Currently, no future X-ray mission is under development to succeed the high-resolution *Chandra* Observatory. Only one mission with sub-arcsec resolution, the X-ray Surveyor,[9] is under conceptual consideration, but it may take more than a decade to launch.

In the near future, several wide-field spectroscopic surveys (e.g., eBOSS, Dawson et al. 2016; PFS, Takada et al. 2014; DESI,[10] Levi et al. 2013; J-PAS,[11] Benitez et al. 2014) will provide galaxy catalogs for higher redshifts, allowing more clusters to be considered, although the yield of high-redshift clusters in the X-ray is not high. Alternatively, a southern redshift survey with BOSS-like depth could easily allow us to double the number of clusters in our analysis, as many of the X-ray clusters in ACCEPT are in the southern hemisphere (see Figure 2). Unfortunately, no such redshift survey to sufficient depth currently exists, although the proposed satellite all-sky redshift survey SPHEREx (Doré et al. 2016) or the planned Euclid survey (Laureijs et al. 2011) could fill in the void. Using simple arguments drawn from a careful analysis of mock simulations, we showed that with a sample of 500 CC and 500 NCC clusters, for which a successor to *Chandra* is crucial, one can improve the measurement presented here from $\Delta b/b = 0.25$ to $\Delta b/b = 0.05$, and rule out (or corroborate) the role of large-scale environment in the creation of the CC/NCC bimodality.

We thank the anonymous referee for useful comments. We acknowledge useful discussions with Megan Donahue, Mark Voit, Keiichi Umetsu, Hironao Miyatake, Ryan Hickox, Andy Goulding, Neal Dalal, and Neta Bahcall. MG is supported by NASA through Einstein Postdoctoral Fellowship Award Number PF-160137 issued by the *Chandra* X-ray Observatory Center, which is operated by the SAO for and on behalf of NASA under contract NAS8-03060. Funding for SDSS-III has been provided by the Alfred P. Sloan Foundation, the Participating Institutions, the National Science Foundation, and the U.S. Department of Energy Office of Science. The SDSS-III website is http://www.sdss3.org/. SDSS-III is managed by the Astrophysical Research Consortium for the Participating Institutions of the SDSS-III Collaboration including the University of Arizona, the Brazilian Participation Group, Brookhaven National Laboratory, Carnegie Mellon University, University of Florida, the French Participation Group, the German Participation Group, Harvard University, the Instituto de Astrofisica de Canarias, the Michigan State/ Notre Dame/JINA Participation Group, The Johns Hopkins University, Lawrence Berkeley National Laboratory, Max Planck Institute for Astrophysics, Max Planck Institute for Extraterrestrial Physics, New Mexico State University, New York University, The Ohio State University, Pennsylvania State University, University of Portsmouth, Princeton University, the Spanish Participation Group, University of Tokyo, University of Utah, Vanderbilt University, University of Virginia, University of Washington, and Yale University. The work reported on in this paper was substantially performed at the TIGRESS high-performance computer center at Princeton University, which is jointly supported by the Princeton Institute for Computational Science and Engineering and the Princeton University Office of Information Technology's Research Computing department.

## Appendix
## Errors from Mock Simulations

When using small samples, resampling (e.g., jackknife) correlation functions can be unrepresentative; we opt to instead use mock simulations to estimate and explore the behavior errors on the bias measurement, as determined from the ratio of CCFs. To that aim, we use the large N-body simulation from the Marenostrum Institut de Ciencies de l'Espai (MICE) collaboration, the MICE Grand Challenge (MICE-GC; Carretero et al. 2015; Crocce et al. 2015; Fosalba et al. 2015a, 2015b; Hoffmann et al. 2015). The galaxy catalog was generated using a hybrid of Halo Occupation Distribution (HOD) and Halo Abundance Matching (HAM) prescriptions to populate Friends of Friends (FoF) dark matter halos from the MICE-GC N-body simulation.[12]

We select both a galaxy and a cluster catalog from the full catalog by limiting both to $0.1 < z < 0.43$, as in our data. We also apply the color selection criteria applied to LOWZ (Reid et al. 2016) using the mock MICE g, r, i magnitudes in the creation of a mock galaxy sample. For the mock cluster samples, we require that the galaxy is central (flag==0). The ACCEPT clusters are massive, spanning $14.4 < \log(M/M_\odot) < 15.1$, so that there are not enough massive simulated clusters in MICE to make a statistically large mock cluster sample from which multiple mocks can be drawn, which still match the observed mass distribution of our CC/NCC clusters. Instead, we construct a cluster sample with a similar mass distribution as that of our clusters, but at a lower mass range, $13.5 < \log(M/M_\odot) < 14.6$. We furthermore divide them equally into two distinct cluster samples, $cl_A$ and $cl_B$, because we are interested in the ratio of CCFs that are of independent clusters. We then randomly select 14 clusters from $cl_A$ and 34 clusters from $cl_B$, the same size as our CC/NCC samples. We repeat this process to produce 50 mock cluster sets. As with the data, we then calculate the CCF of each cluster sample and the mock galaxy catalog.

For the calculation, we produce a random galaxy catalog by drawing a random sample of galaxies in a sphere using VENICE[13] with the same redshift distribution of the MICE-LOWZ mock galaxies (we verify that the construction of the cluster random is not important and its effect cancels out). A simulated "bias" is then constructed from the ratio between any two CCFs, $b_{i,j}(r) = \xi_{i,cl_B,g}(r)/\xi_{j,cl_A,g}(r)$ (2500 ratios in total). The error on the bias is simply the covariance of these 2500 simulated bias measurements. The mean CCFs and their covariance, along with the mean bias and its covariance, are plotted in Figure 7 (left, lower and upper panels, respectively). The covariance of the simulated bias is then used as uncertainty on the bias measured from the data (see Figure 5).

To explore how the errors scale with an increasing number of clusters, we repeat the above test using samples of $N_{cl} = 100$, 250, and 500 clusters in each mock cluster subsample. In Figure 7 (right) we present the CCFs and bias estimated using 500 clusters to demonstrate how we can improve our constraints on the bias. For both small and large samples of clusters, the measured bias is consistent with a null bias ($b = 1$) within the errors, as expected.

---

[9] http://wwwastro.msfc.nasa.gov/xrs/
[10] http://desi.lbl.gov/
[11] http://www.j-pas.org/

[12] http://cosmohub.pic.es/#/catalogs/MICECAT%20v1.0/prebuilt
[13] http://github.com/jcoupon/venice

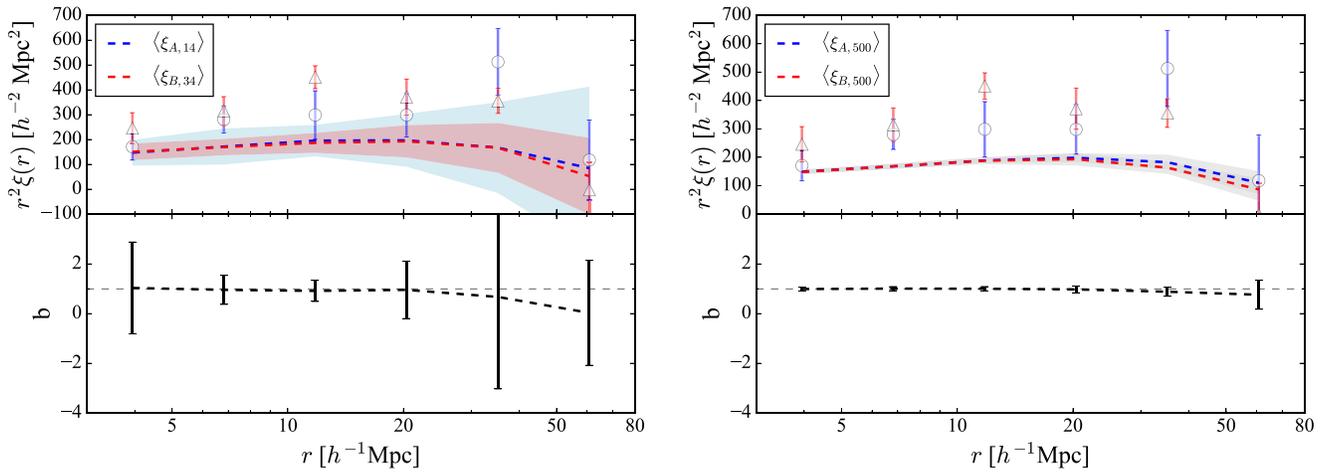

**Figure 7.** Results from 50 mock simulations, using combinations of 14 and 34 cluster subsamples as in our data set (left), and using two subsamples of 500 clusters in each (right). Top panels show the CCF for the data (CC as circles and NCC as triangles) and for the simulations (dashed curve). Shaded regions represent the scatter from the simulations. Bottom panels present the bias as determined from the ratio of the simulated CCFs. Errors on the plots are derived from the diagonal part of the covariance matrix.

As discussed above, we selected mock clusters that are less massive than the real clusters. For this reason, the resulting simulated CCFs (dashed curves with shaded regions) have lower amplitudes than the real CCFs (circles with error bars) in Figure 7. This, however, does not affect the desired quantity (i.e., the bias between the two), as long as they have similar mass distributions.